\documentclass{article}
\usepackage[colorlinks, linkcolor = black, anchorcolor = black, citecolor = blue]{hyperref}
\usepackage[round]{natbib}

\usepackage[ruled,linesnumbered]{algorithm2e}
\usepackage{amsfonts}
\usepackage{amsmath}
\usepackage{subfigure}
\newtheorem{theorem}{Theorem}

\usepackage[utf8]{inputenc}
\usepackage{multirow}
\usepackage{graphicx}
\usepackage{setspace}
\usepackage{authblk}
\title{Differentially Private Confidence Interval for Extrema of Parameters}
\author{Xiaowen Fu$^{a}$\thanks{E-mail address: \it\textbf{xfuak@connect.ust.hk}},
Yang Xiang$^{a,b}$\thanks{E-mail address: \it\textbf{maxiang@ust.hk}},
 Xinzhou Guo$^{a}$\thanks{E-mail address: \it\textbf{xinzhoug@ust.hk}}\\$^{a}$\small\textit{Department of Mathematics, Hong Kong University of Science and Technology,}\\ \small\textit{Clear Water Bay, Kowloon, Hong Kong}\\
 $^{b}$\small\textit{HKUST Shenzhen-Hong Kong
  Collaborative Innovation Research Institute,} \\ \small\textit{Futian, Shenzhen, China}}

\date{}
\begin{document}
\maketitle

\small

\begin{abstract}
This paper aims to construct a valid and efficient confidence interval for the extrema of parameters under privacy protection. The usual statistical inference on the extrema of parameters often suffers from the selection bias issue, and the problem becomes more acute, as in many application scenarios of extrema parameters, we often need to protect the privacy of the data. In this paper, we focus on the exponential family of distributions and propose a privatized parametric bootstrap method to address selection bias in the extrema of parameters problem under the scheme of differential privacy. While the usual privatized parametric bootstrap does not address selection bias appropriately, we prove that with a privatized bias correction term, the proposed parametric bootstrap method can lead to a valid and efficient confidence interval for the extrema of parameters. We illustrate the proposed method with the Gaussian case and regression case and demonstrate the advantages of the proposed method via numerical experiments.
\end{abstract}
\section{Introduction}
	Confidence interval (CI) refers to a range of plausible values for estimates of an unknown parameter of the population. Compared with point estimate, CI not only measures the magnitude of the parameter but also quantifies uncertainty of estimation, and serves as one foundation for statistical inference. CI is widely used in different scientific disciplines. For example, \cite{sandercock2015short} shows that reporting CI has become a standard in medical journals since the late 1980s.  
	
    Extrema parameter refers to the maximum or minimum of some parameters of population and bears practical implications in many real-world problems. It is widely recognized that constructing a CI for the extrema parameter is challenging due to selection bias and simply using sample analogue of the extrema parameter would not lead to valid statistical inference. How to address selection bias and construct a valid and efficient CI for the extrema parameter is an important problem bearing both methodological and practical importance. The problem becomes more acute as in many application scenarios of the extrema parameter, the data are often under privacy protection. 
	
	One important application scenario of extrema parameter is subgroup analysis which aims to uncover and confirm treatment effect heterogeneity within a population. In clinical trials, it is often the case that a treatment is more effective for some patients than others; see for example \cite{mologen2018final}.  When this happens, the extrema parameter can be used to represent the treatment effect of the best subgroup and a valid and efficient CI for the extrema parameter can help researchers better understand the treatment and know where and to what extent the treatment is most useful. Despite the potential benefits, analyzing the best subgroup by directly accessing clinical trial data might face several disincentive issues and in particular, the growing concern of privacy leakage given the sensitive nature of health data as discussed in \cite{xiang2021privacy}.
	
	To protect privacy, several schemes have been proposed. For example, \cite{sweeney2002k} considers forward K-anonymity,\cite{machanavajjhala2007diversity} proposes $l$-diversity and \cite{li2006t} considers $t$-closeness. In this paper, we focus on differential privacy (DP), one widely used scheme proposed by \cite{dwork2006calibrating}. The DP aims to protect privacy by making sure the effect of an arbitrary single substitution in the database is small enough so that the adversarial may not be able to speculate the real data set. As proposed by \cite{dwork2006calibrating}, to achieve DP, we often add some amount of noise to the estimate or data. This will clearly induce additional randomness in statistical inference and the usual CI or bias correction method, assuming the data is public, would not lead to satisfactory results with private data as discussed in \cite{dwork2014algorithmic}
	
	In this paper, we propose a valid and efficient CI for the extrema parameter under the scheme of differential privacy. We focus on the exponential family of distributions and develop a privatized parametric bootstrap approach to address selection bias in the extrema parameter under the scheme of differential privacy. The proposed method is easy-to-implement, efficient, and can be adapted to different practical scenarios.

\subsection{Related Literature}
    \textbf{De-Biased CI for Extrema Parameter} It is well recognized that without appropriate adjustment, CI for the extrema parameter will suffer from selection bias; see \cite{thomas2017comparing} and \cite{magnusson2013group}.  To construct a de-biased CI for the extrema parameter, some attempts have been made. \cite{fuentes2018confidence} and \cite{hall2010bootstrap} propose valid CI based on simultaneous controls so the resulting CIs tend to be conservative and are not efficient. \cite{rosenkranz2016exploratory} and \cite{stallard2008estimation} have considered some ad-hoc methods to adjust selection bias, but those methods lack theoretical justifications.   \cite{bornkamp2017model} and \cite{woody2022optimal} consider Bayesian approach, which is clearly model-dependent and often lacks frequentist interpretation. Recently, several bootstrap-based CI for extrema parameter have been proposed in \cite{guo2021inference} and \cite{guo2022assessing} among others. While these bootstrap-based CI for extrema the parameter are efficient and well-justified, they are designed for the public data and are not directly applicable under the DP scheme.

    \textbf{Differentially Private Inference} Statistical inference under DP has been considered for several scenarios. For example,  \cite{dimitrakakis2014robust}, \cite{dimitrakakis2017differential}, \cite{zhang2016differential} study DP of Bayesian inference, \cite{rogers2016max}, \cite{balle2020hypothesis}, \cite{gaboardi2016differentially} do hypothesis testing under DP, and \cite{rinott2018confidentiality} study DP in frequency tables. There are some works on constructing CI with DP. For example, \cite{karwa2017finite} proposes a private algorithm to estimate a range for the data and derives a private CI. \cite{du2020differentially} proposes private simulation and quantile methods for estimating mean and variance. \cite{wang2018differentially} proposes the method to construct CI under differential privacy for empirical risk estimation which can be applied to logistic regression and support vector machines (SVM).
    
    \textbf{Differentially Private Bootstrap} To correct bias, bootstrap procedures are often adopted. Under DP, \cite{covington2021unbiased} develops bag of little bootstraps (BLB) to privately estimate sampling distribution of parameters, \cite{chadha2021private} proposes a private confidence sets with bootstrap, \cite{brawner2018bootstrap} uses the bootstrap with DP to estimate standard errors “for free”, \cite{dunsche2022multivariate} presents a test for multivariate mean comparisons under pure-DP with bootstrap, and \cite{ferrando2022parametric} proposes a method to construct confidence intervals with differentially private parametric bootstrap. However, the existing bootstrap procedures under DP fail to address selection bias appropriately and can not deliver valid CI for the extrema parameter.
    
    \textbf{Summary} As far as we know, under DP, the usual bootstrap cannot be directly used to address selection bias and de-biased CI for the extrema parameter is still lacking, and we aim to bridge this gap in this paper.  

\subsection{Contribution}
The main contribution of our work can be summarized as follows: (1) propose a privatized parametric bootstrap procedure to address selection bias in the extrema parameter under differential privacy; (2) account for randomness induced by noise term in constructing CI with private data; (3) implement our proposed method with Gaussian case and regression case which have broad applications in practice; and (4) propose strategies to avoid wasting privacy budget on nuisance parameter.

\section{Preliminaries}
\subsection{Extrema Parameter and Selection Bias} Let $\beta=(\beta_1,\dots,\beta_k) \in \mathbb{R}^k$ denote some parameters of interest of population and $\hat{\beta}_j$ denote an estimate for $\beta_j$ for $j=1,\cdots,k$ from data sets $\mathcal{X}$.  W.L.O.G., our goal is to construct a lower confidence limit for the maximum $\beta_{\max} = \max\limits_{j = 1,\cdots,k} \beta_j$. 

Due to selection bias, the sample analogue of $\beta_{\max}$, $\hat{\beta}_{\max}=\max\limits_{j = 1,\cdots,k} \hat{\beta}_j$, is biased even when $\hat{\beta}_j$ is consistent to $\beta_j$, and CI simply relying on $\hat{\beta}_{\max}$ would not be valid; see \cite{nadarajah2008exact} for theoretical derivations. 

\subsection{Differential Privacy} 

Differential privacy is a scheme for privacy protection that if changing an individual data in the data set does not cause much change in the outcome, the adversarial may not be able to speculate the real data set.

\textbf{Definition 1} (Differential privacy,  \cite{dwork2006calibrating}) A mechanism $\mathcal{A}$ is said to satisfy $\varepsilon$ - differential privacy ($\varepsilon$-DP) if for all pairs $x,x^{'} \in \mathcal{X}$ which differ in only one entry, and for any outcome $O \subseteq \text{range}(\mathcal{A})$, we have 
\begin{displaymath}
    |\ln(\frac{Pr(\mathcal{A}(x)\in O)}{Pr(\mathcal{A}(x^{'})\in O)})| \leq \varepsilon
\end{displaymath}

Under DP, $\varepsilon$ is a parameter to control privacy leakage and is called privacy budget. A smaller $\varepsilon$ indicates better privacy protection at the potential cost of accuracy.  To achieve differential privacy, we often need to add some amount of noise to the data and the amount is often determined by the sensitivity defined as follows \cite{dwork2006calibrating}.

\textbf{Definition 2} (Sensitivity) The sensitivity of a function $\Gamma$ is the smallest number $S(\Gamma)$ such that for all $x,x^{'}\in \mathcal{X}$ which differ in a single entry,
\begin{displaymath}
\setlength{\abovedisplayskip}{3pt}
\setlength{\belowdisplayskip}{3pt}
    ||\Gamma(x)-\Gamma(x^{'})||_1 \leq S(\Gamma)
\end{displaymath}

For a random algorithm $\Gamma$, to achieve $\varepsilon$-DP, we often consider the Laplace mechanism. 

\textbf{Definition 3} (Laplace mechanism)
For all function $\Gamma$ that maps data sets to  $\mathbb{R}^d$, $\Gamma(\mathbf{x}) + w$ is $\varepsilon$-DP, where $w=\{w_k\}_{k=1}^d$ is the added Laplacian noise with entry $w_k \sim \text{\text{Lap}}(S(\Gamma)/\varepsilon)$, and $\text{Lap}$ denotes a zero-mean Laplacian distribution with scale $S(\Gamma)/\varepsilon$.

The Laplace mechanism introduces additional randomness to protect privacy, which usually brings damage to accuracy and efficiency in estimation or inference and we need to appropriately account for it. 

Differential privacy has the composition properties as follows \cite{zhao2017composition}, which we use later for privacy budget allocation for parameters and privacy guarantee in cross-validation.

\textbf{Sequential Composition Theorem} Let $M_i$ each provide $\varepsilon_i$-DP, then the sequence of $M_i(X)$ provides ($\sum_i \varepsilon_i$)-DP. 

\textbf{Parallel Composition Theorem} Let $M_i$ each provide $\varepsilon$-DP. Let $D_i$ be arbitrary disjoint subsets of the input domain $D$. The sequence $M_i(X\cap D_i)$ provides $\varepsilon$-DP.

\subsection{Parametric Bootstrap}
Bootstrap is a resampling procedure using random samples with replacement to mimic the sampling process and is often adopted for statistical inference and to correct bias. However, drawing samples with replacement might incur a leakage of privacy. To avoid accessing original data many times in bootstrapping, we might consider parametric bootstrap assuming that we know the distribution family of data sets denoted by $X \sim \mathcal{M}(\alpha)$, where $\alpha$ is the natural parameter in the distribution. Let $\theta$ denote the parameter of interest, a parametric bootstrap CI for $\alpha$ is shown in Algorithm \ref{alg1}. 

\begin{algorithm}
	\SetAlgoLined
	\KwData{$\mathcal{X}$}
	\KwResult{CI of $\theta$}
	Derive the estimate of the natural parameter $\hat{\alpha}$ based on $\mathcal{X}$\;
	\While{$b=1:B$}{
		Generate bootstrap sample based on estimated distribution $\mathcal{X}^{*} \sim \mathcal{M}(\hat{\alpha})$\;
		Estimate the target statistics based on bootstrap sample $\theta^{*}_b = \theta^{*}(\mathcal{X}^{*})$
	}
	Derive CI based on $\theta^{*}_1,\cdots,\theta^{*}_B$.
	
\caption{Parametric bootstrap CI for $\theta$}
\label{alg1}
\end{algorithm}

\section{Differentially Private Confidence Interval for Extrema Parameter}
In this section, we introduce the framework of deferentially private CI for extrema parameter. The framework built on exponential family could be naturally extended to other parametric families of distribution with focus on the exponential family distribution.  

\subsection{Exponential Family Distribution}
For a sample $x_i$ from the data set $\mathcal{X}=\{x_i\}_{i=1}^n$, a family of distribution is called the exponential family if the density function is
\begin{displaymath}
\setlength{\abovedisplayskip}{3pt}
\setlength{\belowdisplayskip}{3pt}
p(x_i;\alpha) = h(x_i)e^{\alpha^T T(x_i)-A(\alpha)}
\end{displaymath}
where $h$ is a base function, $\alpha$ is the natural parameter, $T$ is the sufficient statistics function, and $A(\alpha)$ is the log-partition function. Exponential family distribution includes many common distributions.

In practical applications, the parameter of our interest $\beta \in \mathbb{R}^{k_1}$ can be viewed as a function of $\alpha$; i.e. $\beta = f(\alpha)$. With appropriate nuisance parameter $\gamma \in \mathbb{R}^{k_2}$, there exists a partition of $\alpha=(\alpha_1,\alpha_2)$ and an a 1-1 mapping $\mathbf{f}$: $(\alpha_1,\alpha_2) = \mathbf{f}(\beta,\gamma)$ where $\alpha_2$ only depends on the nuisance parameter $\gamma$ and

\begin{equation}
\setlength{\abovedisplayskip}{3pt}
\setlength{\belowdisplayskip}{3pt}
    \begin{cases}
\alpha_1 = f_1(\beta,\gamma)\\
\alpha_2 = f_2(\gamma).
\label{equa1}
\end{cases}
\end{equation}
$\alpha_1$ can be generated by gathering all the terms including $\beta$, and $\alpha_2$ is then easy to determine. With this reparameterization. the exponential family distribution can be rewritten as 
\begin{equation}
\setlength{\abovedisplayskip}{3pt}
\setlength{\belowdisplayskip}{3pt}
p(x_i;\alpha_1,\alpha_2) = h(x_i)e^{\alpha_1^T T_1(x_i)+\alpha_2^T T_2(x_i)-A(\alpha_1,\alpha_2)} \label{equa2}
\end{equation}
where $(T_1(x_i),T_2(x_i))=T(x_i)$. Our goal is to construct private CI for $\beta_{\max} = \max\limits_{j =1,\cdots,k_1}\beta_j$. 

Considering data sets $\mathcal{X}=\{x_i\}_{i=1}^n$ and plugging Eq. \ref{equa1} into Eq. \ref{equa2}, the log-likelihood is 
\begin{align}
\setlength{\abovedisplayskip}{3pt}
\setlength{\belowdisplayskip}{3pt}
    \ln p(\mathcal{X};\beta,\gamma) &= \sum \ln h(x_i) + f_1(\beta,\gamma) \sum T_1(x_i) \nonumber \\
&+f_2(\gamma) \sum T_2(x_i) - n A(\beta,\gamma) \nonumber
\end{align}
where $A(\beta,\gamma) = A(f_1^{-1}(\beta,\gamma),f_2^{-1}(\gamma))$ and $\sum$ denotes the simplified symbol for $\sum_{i=1}^n$ throughout the paper. Then, the MLE estimate is a solution for Eq.  \ref{equa3}.

\begin{equation}
\setlength{\abovedisplayskip}{3pt}
\setlength{\belowdisplayskip}{3pt}
\begin{aligned}
    \frac{\partial \ln p(X;\beta,\gamma)}{\partial \beta} &= \frac{\partial  f_1(\beta,\gamma)}{\partial \beta}\sum T_1(x_i) - n \frac{\partial A}{\partial \beta} = 0\\
    \frac{\partial \ln p(X;\beta,\gamma)}{\partial \gamma} &= \frac{\partial  f_1(\beta,\gamma)}{\partial \gamma}\sum T_1(x_i) \\
    &+\frac{\partial f_2 (\gamma)}{\partial\gamma}\sum T_2(x_i) - n \frac{\partial A}{\partial \gamma} = 0 \label{equa3}
    \end{aligned}
\end{equation}

We write out the solution in preparation for the partially privatized case discussed in Section 5. The solution can be written as \begin{equation}
\setlength{\abovedisplayskip}{3pt}
\setlength{\belowdisplayskip}{3pt}
    \begin{cases}
        \hat{\beta} = g_1(\sum T_1(x_i),\sum T_2(x_i)) \\
        \hat{\gamma} = g_2(\sum T_1(x_i),\sum T_2(x_i)) \label{equa4}
    \end{cases}
\end{equation}
where $g_1,g_2$ are functions of sufficient statistics to stand for the MLE of $\beta,\gamma$. 

\subsection{Privatized Parametric Bootstrap CI}
To construct differentially private CI for $\beta_{\max}$, we propose a privatized parametric bootstrap algorithm to address selection bias in the extrema parameter. There are three key elements in the algorithm: (1) privatized point estimate; (2) privatized parametric bootstrap and (3) privatized bias-correction term. 

\textbf{Privatized Point Estimate}: To achieve differential privacy in point estimation, following Eq. \eqref{equa4}, we add Laplacian noises to sufficient statistics in Eq. \ref{equa4} in Step 2 as follows
\begin{equation}
\setlength{\abovedisplayskip}{3pt}
\setlength{\belowdisplayskip}{3pt}
    \begin{cases}
        \hat{\beta}^{priv} = g_1(\sum T_1(x_i)+w_1,\sum T_2(x_i)+w_2) \\
        \hat{\gamma}^{priv} = g_2(\sum T_1(x_i)+w_1,\sum T_2(x_i)+w_2) \label{equa5}
    \end{cases}
\end{equation}
where $w_i\sim Lap(\triangle_i/\varepsilon_i)(i = 1,2)$ is the Laplacian noise. $\triangle_i$ is the sensitivity of sufficient statistics $\sum T_i(x)$, and $\varepsilon_i$ is the privacy budget allocated to it. According to sequential composition theory, the framework is $(\varepsilon_1+\varepsilon_2)$-DP.

\textbf{Privatized Parametric Bootstrap}: To avoid accessing data repeatedly, parametric bootstrap is adopted here. In specific, following the idea in \cite{ferrando2022parametric}, we generate bootstrap estimate from the estimated model based on $\hat{\beta}^{priv}$ and $\hat{\gamma}^{priv}$. Because we can at most infer the estimated model (i.e. $\hat{\beta}^{priv}$ and $\hat{\gamma}^{priv}$) from the bootstrap sample,  by parallel composition theory, the framework remains $(\varepsilon_1,\varepsilon_2)$-DP after bootstrap. To account for randomness induced in Laplace scheme, we add a Laplace noise and calculate bootstrap estimate $\hat{\beta}_{j}^{*,priv}$ in Step 6. 

\textbf{Privatized Bias-Correction Term}: It is well known that the usual bootstrap can not address selection bias; see \cite{bornkamp2017model}. Following the idea of \cite{guo2021inference}, we consider a modified bootstrap estimate, $\hat{\beta}_{j,modified}^{*,priv}$, to correct selection bias,
    \begin{displaymath}
    \setlength{\abovedisplayskip}{3pt}
\setlength{\belowdisplayskip}{3pt}
    \hat{\beta}_{j,modified}^{*,priv} = \hat{\beta}_{j}^{*,priv} + d_j,j = 1,\cdots,k_1,
    \end{displaymath}
    where $d_j$ is the distance of privatized estimate $\hat{\beta}_j^{priv}$ and its extrema based on the original data sets
    \begin{displaymath}
    \setlength{\abovedisplayskip}{3pt}
\setlength{\belowdisplayskip}{3pt}
    d_j = (1-n^{r-0.5})(\hat{\beta}_{\max}^{priv}-\hat{\beta}_j^{priv}),j = 1,\cdots,k_1.
    \end{displaymath}
    where $n$ is the size of the total population ,and $r \in (0,0.5)$ is a tuning parameter, which determines the adjustments to the estimate. With a smaller $r$, the adjustment gets stronger, and the coverage probability gets better with the sacrifice of the efficiency of confident limit. 

The framework remains $(\varepsilon_1,\varepsilon_2)$-DP after correction. Take the lower confidence limit as an example, the proposed algorithm is summarized in Algorithm \ref{alg2}.

\begin{algorithm}
	\SetAlgoLined
	\KwData{$x_1,\cdots,x_n$}
	\KwResult{Privatized lower confidence limit of $\beta_{\max}$}
	Add Laplacian noise to sufficient statistics and calculate the privatized MLE $\hat{\beta}^{priv},\hat{\gamma}^{priv}$ based on Eq. \ref{equa5}\;
	Estimate privatized extrema $\hat{\beta}_{\max}^{priv} = \max\limits_{j = 1,\cdots,k_1}\{\hat{\beta}_j^{priv}\}$\;
	Calculate bias-correction term $d_j = (1-n^{r-0.5})(\hat{\beta}_{\max}^{priv}-\hat{\beta}_j^{priv}),j=1,\cdots,k_1$\;
	\While{$b=1:B$}{
		Generate bootstrap sample based on exponential family (\ref{equa2}) and parameter transformation (\ref{equa1}) $x_1^{*},\cdots,x_n^{*} \sim p(f_1(\hat{\beta}^{priv},\hat{\gamma}),f_2(\hat{\gamma}^{priv}))$\;
		Add Laplacian noise to sufficient statistics and calculate the privatized MLE $\hat{\beta}^{*,priv}$ based on  (\ref{equa5})\;
		Calculate $T^{*,priv} = \sqrt{n} (\max\{\hat{\beta}_j^{*,priv}+d_j - \hat{\beta}_{\max}^{priv}\})$
	}
	Let $c_\alpha = quantile (T^{*,priv},1-\alpha)$. The level $1-\alpha$ lower confidence limit is $L^{priv}=\hat{\beta}_{\max}^{priv}-c_\alpha/\sqrt{n}$ 
\caption{Privatized parametric bootstrap inference on extrema problems with bias-correction}
\label{alg2}
\end{algorithm}

\subsection{Cross-Validation}
To choose $r$, we suggest a data-adaptive cross-validated method. To start with,  we consider a bias-reduced estimate $\hat{\beta}_{\max,reduced}^{priv}$ as follows:
	\begin{equation}
	\setlength{\abovedisplayskip}{3pt}
\setlength{\belowdisplayskip}{3pt}
		\hat{\beta}_{\max,reduced}^{priv} = \hat{\beta}_{\max}^{priv}-E^{*}[\beta_{\max,modified}^{priv,*}-\hat{\beta}_{\max}^{priv}], \label{equa6}
	\end{equation}
	where $E^{*}$ denotes expectation under bootstrap distribution.
	
	 The idea of cross-validation is to choose $r$ to minimize the mean square error between $\hat{\beta}_{\max,reduced}^{priv}(r)$ and $\beta_{\max}^{priv}$. Let $A =\{r_1,\cdots,r_m\}$ denote a set of possible tuning parameters in the range of $(0,0.5)$ with $r_1<\cdots<r_m$ and $m$ is a finite integer. The following algorithm can be used to choose $r\in A$ under differential privacy based on our framework.
	
	For the $j$-fold in cross validation, we denote $\hat{\beta}_{j}^{priv} = \{\hat{\beta}_{j,i}^{priv}\}_{i=1}^{k_1}$ with $\hat{\beta}_{j,i}^{priv}$ as the $i$-th item of $\hat{\beta}_{j}^{priv}$. According to parallel composition theory, the framework reserves the privacy budget as the one in Algorithm  \ref{alg2}.
	
	\begin{algorithm}[h]
		\SetAlgoLined
		\KwData{$x_1,\cdots,x_n$}
		\KwResult{Optimal choice of tuning parameter $r$}
		Randomly partition the data into $v$ (approximately) equalsized subsamples\;
		\For{$l = 1,\cdots,m$}{
			\For{$j = 1,\cdots,v$}{
				Use the $j$th subsample as the reference data and the rest as the training data\;
				Use the training data to obtain the bias-reduced estimate with DP via (\ref{equa6}) $\hat{\beta}_{\max,reduced,j}^{priv}(r_l)$, with $r_l$ as the tuning prarmeter\;
				Use the reference data to estimate  $\hat{\beta}_{j}^{priv}$\;
				\For{$i = 1,\cdots,k_1$}{
				Calculate the standard error $\hat{\sigma}_{j,i}^{priv}$ for $\hat{\beta}_{j,i}^{priv}$ \;
					Calculate accuracy of each choice $h_{j,i}^{priv}(r_l)=(\hat{\beta}_{\max,reduced,j}^{priv}(r_l)-\hat{\beta}_{j,i}^{priv})^2-(\hat{\sigma}_{j,i}^{priv})^2$
				}
			}
		}
	The tuning parameter is chosen to be $argmin_{r_l}\{\min\limits_{i\in [k]}[\sum_{j =1}^{j = v}h_{j,i}^{priv}(r_l)]/v\}$

		\caption{Cross-validated choice of tuning parameter $r$}
	\end{algorithm}

\section{Application}
In this section, we apply our proposed framework to two important scenarios: (1) multivariate Gaussian and (2) linear regression. For simplicity, we only discuss the implementation of the privatized point estimation and privatized parametric bootstrap as the detailed algorithm naturally follows from the framework in Algorithm \ref{alg1}. 

\subsection{Multivariate Gaussian Case}
Consider a $k$-dimensional multivariate Gaussian where $\mathbf{x}_i\sim N(\mu,\Sigma)$ for $i=1,\dots,n$,  $\beta = \mu \in \mathbb{R}^{k}$ is the parameter of interest and $\gamma = \Sigma \in \mathbb{R}^{k\times k}$ is an nuisance parameter. Then, $\beta_{\max}$ represents the largest population mean and often bears practical implications, such as the best subgroup effect in clinical studies. For $\mathbf{x}_i$, the density function is
\begin{displaymath}
\setlength{\abovedisplayskip}{3pt}
\setlength{\belowdisplayskip}{3pt}
p(\mathbf{x}_i;\mu,\Sigma) = \frac{1}{(2\pi)^{\frac{k}{2}}|\Sigma|^\frac{1}{2}}e^{-\frac{(\mathbf{x}_i-\mu)^T\Sigma^{-1}(\mathbf{x}_i-\mu)}{2}}
\end{displaymath}
with two sufficient statistics: $T_1(\mathbf{x}_i) =\mathbf{x}_i, T_2(\mathbf{x}_i) = \mathbf{x}_i\mathbf{x}_i^T$. Given data sets  $\mathcal{X} = \{\mathbf{x}_i\}_{i=1}^n$, the log-likelihood is
\begin{align}
\setlength{\abovedisplayskip}{3pt}
\setlength{\belowdisplayskip}{3pt}
\setlength{\abovedisplayskip}{3pt}
\setlength{\belowdisplayskip}{3pt}
\ln p (\mathcal{X};\mu,\Sigma) &= -\frac{nk\ln(2\pi)}{2}-\frac{n\ln(|\Sigma|)}{2} \nonumber\\
&-\frac{1}{2}\sum_{i = 1}^n(\mathbf{x}_i-\mu)^T\Sigma^{-1}(\mathbf{x}_i-\mu) \nonumber, 
\end{align}
and the privatized point estimate by MLE is
\begin{equation}
\left\{\begin{aligned}
\hat{\mu}^{priv} &= \frac{1}{n}(\sum\limits_{i = 1}^n \mathbf{x}_i +w_1)  \\
\hat{\Sigma}^{priv} &= \frac{1}{n-1} (\sum\limits_{i = 1}^n
\mathbf{x}_i\mathbf{x}_i^T +w_2) \\ &-\frac{1}{n(n-1)} (\sum\limits_{i = 1}^n\mathbf{x}_i+w_1)(\sum\limits_{i = 1}^n\mathbf{x}_i+w_1)^T
\end{aligned}\right.
\label{equa7}
\end{equation}
where $w_i\sim Lap(\triangle_i/\varepsilon_i)(i = 1,2)$ is the Laplacian noise. $\triangle_i$ is the sensitivity of sufficient statistics $\sum T_i(x)$, and $\varepsilon_i$ is the privacy budget allocated to it.

For the privatized bootstrap, we generate $\mathbf{x}_1^{*},\cdots,\mathbf{x}_n^{*} \sim N(\hat{\mu}^{priv},\hat{\Sigma}^{priv})$ and the estimate
\begin{equation}
    \hat{\mu}^{*,priv} = \frac{1}{n}(\sum\limits_{i = 1}^n \mathbf{x}_i^{*} +w_1^{*})\label{equa8}
\end{equation}
where $w_1^{*}$ is the Laplacian noise generated by the same distribution of $w_1$ to account for the additional randomness in privacy protection. With $\hat{\mu}^{priv}$ and $\hat{\mu}^{*,priv}$, we can proceed following Algorithm \ref{alg2}.

\subsection{Linear Regression}
We consider the linear regression case. While inspired by exponential family, some modifications are adopted to better fit the protection of privacy in parametric bootstrap detailed later. Consider a linear model $y_i = \mathbf{x}_i^T\beta+e_i$, where $e_i \sim N(0,\sigma^2)$, $\beta,\mathbf{x}_i \in \mathbb{R}^k,y_i \in \mathbb{R}$ for $i=1,\dots,n$. Then, $\beta_{\max}$ represents the largest regression coefficient and often bears practical implications, such as the strongest signal in genetic association studies. The density function is
\begin{displaymath}
p(\mathbf{x}_i,y_i;\beta,\sigma^2) = \frac{1}{\sqrt{2\pi \sigma^2}}e^{-\frac{(y_i-\mathbf{x}_i^T\beta)^2}{2\sigma^2}}
\end{displaymath}
 Let $X \in \mathbb{R}^{n\times k}$ denote the matrix with the $i^{th}$ row equal to $\mathbf{x}_i^T$, and $\mathbf{y},\mathbf{e} \in \mathbb{R}^n$ be the vectors with the $i^{th}$ entries $y_i$ and $e_i$, respectively. Then the linear regression problem becomes $\mathbf{y} = X\beta+\mathbf{e}$. There are  The log-likelihood is
 \begin{displaymath}
 \setlength{\abovedisplayskip}{3pt}
\setlength{\belowdisplayskip}{3pt}
 \begin{aligned}
 \ln P(X,\mathbf{y};\beta,\sigma^2) &= -\frac{\mathbf{y}^T\mathbf{y}-2\beta^TX^T \mathbf{y}+\beta^T X^T X \beta}{2\sigma^2} \\
 &-\frac{n}{2}\ln(\sigma^2),
 \end{aligned}
 \end{displaymath}
 
 and the privatized point estimate for $\beta$ by MLE is \begin{equation}
 \setlength{\abovedisplayskip}{3pt}
\setlength{\belowdisplayskip}{3pt}
     \hat{\beta}^{priv} = (X^TX+w_1)^{-1}(X^T\mathbf{y}+w_2) . \label{equa9}
 \end{equation}
 where $w_i \sim \text{Lap}(\triangle_i/\varepsilon_i)(i=1,2)$ is the Laplacian noise, $\triangle_i$ is the sensitivity of sufficient statistics $\sum T_1 = X^T X$, and $\sum T_2 = X^T \mathbf{y}$. Following \cite{ferrando2022parametric}, we adopt a bias-corrected estimate for $\sigma^2$ with additional Laplacian noise. 
 
 \begin{equation}
 \setlength{\abovedisplayskip}{3pt}
\setlength{\belowdisplayskip}{3pt}
    \hat{\sigma}^{2,priv} = \frac{1}{n-k}\left[\sum\limits_{i=1}^n (y_i-\mathbf{x}_i^T \hat{\beta}^{priv})^2\right] +w_3,\label{equa10}
 \end{equation}

where $w_3 \sim \text{Lap}(\triangle_3/\epsilon_3)$ is additional the Laplacian noise, and $\triangle_3$ is the sensitivity of the variance. This scheme is also beneficial to the partially privatized settings to be discussed in the next section.

If we use the estimated model $\mathbf{y}=X\hat{\beta}^{priv}+\mathbf{e}$ to generate bootstrap sample,  we need to access the original data $X$ many times and sacrifice privacy budget to protect $X$. \cite{ferrando2020general} suggests we rewrite the privatized MLE $\hat{\beta}^{priv}$ in Eq. \ref{equa9} and consider the following bootstrap estimate for $\beta$
\begin{align}
\setlength{\abovedisplayskip}{3pt}
\setlength{\belowdisplayskip}{3pt}
	\sqrt{n}\hat{\beta}^{*,priv} &= \sqrt{n}(S^{priv}+\frac{1}{n}w_1^{*})^{-1}S^{priv}\hat{\beta}^{priv} \nonumber \\ &+ (S^{priv}+\frac{1}{n}w_1^{*})^{-1}(C^{*,priv}+\frac{1}{\sqrt{n}}w_2^*) \label{equa11}
\end{align}
where $S^{priv}= \frac{1}{n}X^TX+\frac{1}{n}w_1$, and we generate $C^{*,priv} \sim N(0,\hat{\sigma}^{2,priv} S^{priv})$. $w_1,w_2$ are corresponding Laplacian noises, and $w_1^{*},w_2^{*}$ are the Laplacian noises generated by the same distribution of $w_1,w_2$ to account for the addition randomness in privacy protection. With $\hat{\beta}^{priv}$ and $\hat{\beta}^{*,priv}$, we can proceed following Algorithm \ref{alg2}.

\section{Partially Private Method}
In this section, we introduce a strategy to save privacy budget when the calculation and bootstrapping of parameter of interest $\beta$ only depends on part of sufficient statistics $T$. We discuss the implementation of the strategy in two applications for multivariate Gaussian and linear regression. 

\subsection{General Privacy Budget Reduction} 

In some practical applications, the release of the estimate of parameter of interest $\beta$ might only depends on part of sufficient statistics as stated in Theorem \ref{thm4}.   

\begin{theorem}
In the framework of exponential family shown in section 3.1, if $\frac{\partial A}{\partial f_1} = l(\beta)$, $l$ is some function, then the MLE has the form,
\begin{equation}
\setlength{\abovedisplayskip}{3pt}
\setlength{\belowdisplayskip}{3pt}
    \begin{cases}
        \hat{\beta} = g_1(\sum T_1(x_i)) \\
        \hat{\gamma} = g_2(\sum T_1(x_i),\sum T_2(x_i)) \label{equa12}.
    \end{cases}
\end{equation}\label{thm4}
\end{theorem}

From Eq. \ref{equa12}, we see that $T_2(x)$ has nothing to do with the release of $\hat{\beta}$, which implies that for estimation, we may save the privacy budget without adding Laplacian noise to $T_2(x)$ or in specific. Let
\begin{equation}
\setlength{\abovedisplayskip}{3pt}
\setlength{\belowdisplayskip}{3pt}
    \begin{cases}
        \hat{\beta}^{priv} = g_1(\sum T_1(x_i)+w_1) \\
        \hat{\gamma} = g_2(\sum T_1(x_i)+w_1,\sum T_2(x_i)).
    \end{cases}\label{equa13}
\end{equation}
We obtain an $\varepsilon_1$-DP estimation instead of the $(\varepsilon_1+\varepsilon_2)$-DP with Eq.  \ref{equa13}. Furthermore, if the release of $\hat{\beta}^*$ has nothing to do with $\hat{\gamma}$ or only depends on part of $\hat{\gamma}$, we may only need to add noise to the relevant part in estimating $\gamma$ and save the privacy budget in Algorithm \ref{alg2}. We state the partially privatized scheme of Gaussian case and regression case, and more details are included in Appendix.

\subsection{Partially Privatized Multivariate Gaussian Case} 

In applications, we might be only interested in some of the population means. Suppose \begin{equation}
\setlength{\abovedisplayskip}{3pt}
\setlength{\belowdisplayskip}{3pt}
\mathbf{x}_i= 
    \left(
    \begin{array}{c}
         \mathbf{x}_i^1\\
         \mathbf{x}_i^2
    \end{array}\right) \sim 
    \left(
    \left(\begin{array}{c}
         \mu_1\\
         \mu_2
    \end{array}
    \right),
    \left(
    \begin{array}{cc}
         \Sigma_{11}&  \Sigma_{12}\\
         \Sigma_{21}&  \Sigma_{22}
    \end{array}
    \right)
    \right)
\end{equation}
where $\mathbf{x}_i^1 \in \mathbb{R}^{n_1}$, $\mathbf{x}_i^2 \in \mathbb{R}^{n_2}$. Then $\beta = \mu_1$ is the parameter of interest, and $\gamma = (\mu_2,\Sigma)$ is the nuisance parameter, where $\Sigma = \left(
    \begin{array}{cc}
         \Sigma_{11}&  \Sigma_{12}\\
         \Sigma_{21}&  \Sigma_{22}
    \end{array}
    \right)$. 
We can check that in this multivariate Gaussian case, $\frac{\partial A}{\partial f_1}=\frac{1}{\mu_1}$, which satisfies the condition in Theorem \ref{thm4}. Note that in parametric bootstrap, we only need to estimate $\mu$ and $\Sigma_{11}$ instead of all, we can modify Eq. \ref{equa7} and derive the partially privatized MLE estimate by only adding Laplacian noise $w_1,w_2$ to sufficient statistics $\sum\limits_{i=1}^n \mathbf{x}_i^1,\sum\limits_{i=1}^n \mathbf{x}_i^1(\mathbf{x}_i^1)^T$.

We estimate
\begin{displaymath}
\setlength{\abovedisplayskip}{3pt}
\setlength{\belowdisplayskip}{3pt}
\hat{\mu}_1^{*,priv} = \frac{1}{n_1}(\sum\limits_{i = 1}^{n_1} \mathbf{x}_i^{*} +w_1^{*})
\end{displaymath}
where $w_1^{*}$ is the Laplacian noise generated by the same distribution of $w_1$. Then we calculate the $T^{*,priv}$ and the privatized lower confident limit as in Algorithm \ref{alg2}. As for the bias-selection correction parts in step 3 and step 7 in Algorithm \ref{alg2}, we plug in partial sample size $n_1$.

With this modification, we can save the privacy budget for $\mu_2,\Sigma_{12},\Sigma_{21}$ and $\Sigma_{22}$.

\subsection{Linear Regression with Nuisance Parameters}
In many applications, we often consider a linear regression model $y_i = \mathbf{z}_i^T\beta +\mathbf{x}_i^T\gamma +e_i$, $i = 1,\cdots,n$. Take subgroup analysis as an example, $\mathbf{z}_i \in \mathbb{R}^{k_1}$ can be the interaction terms between subgroup indicators and treatment indicator, $\mathbf{x}_i \in \mathbb{R}^{k_2}$ can be pre-trement covariates, and $y_i$ is the response vector, $e_i \sim N(0,\sigma^2),i.i.d.$ is the white noise; see \cite{imai2013estimating}. Then, $\beta \in \mathbb{R}^{k_1}$ is the parameter of interest and $\beta_{\max}$ represents the best subgroup effect, and $\gamma \in \mathbb{R}^{k_2}$ is the nuisance parameter. 

Let $Z \in \mathbb{R}^{n\times k_1}$ denote the matrix with the $i^{th}$ row equal to $\mathbf{z}_i^T$, $X \in \mathbb{R}^{n\times k_2}$ denote the matrix with the $i^{th}$ row equal to $\mathbf{x}_i^T$ and $\mathbf{y},\mathbf{e} \in \mathbb{R}^n$ denote the vectors with the $i^{th}$ entries $y_i$ and $e_i$, respectively.

In some real problems such as randomized trails, we have $Z^T X = 0$. We can check that $A(\beta,\gamma,\sigma^2) = \frac{1}{2}\ln \sigma^2$. Thus $\frac{\partial A}{\partial \beta} = 0$. The condition of Theorem \ref{thm4} is satisfied. We then follow the scheme of Eq. \ref{equa9} to construct privatized MLE for $\beta$ by only adding Laplacian noises $w_1,w_2$ to 2 sufficient statistics $\sum T_1 =Z^T Z$ and $\sum T_3 = Z^T \mathbf{y}$ related to $\beta$. 

We adopt a bias-corrected estimate $\hat{\sigma}^{2,priv}$ for $\sigma^2$ with additional Laplacian noise following the form of Eq. \ref{equa10} by plugging $\hat{\beta}^{priv},\hat{\gamma}$.

Similarly, we form a bootstrap estimate $\hat{\beta}^{*,priv}$ that follows the idea of Eq. \ref{equa11}:
\begin{align}
\setlength{\abovedisplayskip}{3pt}
\setlength{\belowdisplayskip}{3pt}
	\sqrt{n}\hat{\beta}^{*,priv} &= \sqrt{n}(S^{priv}+\frac{1}{n}w_1^{*})^{-1}S^{priv}\hat{\beta}^{priv} \nonumber \\ &+ (S^{priv}+\frac{1}{n}w_1^{*})^{-1}(C^{*,priv}+\frac{1}{\sqrt{n}}w_2^*), \label{equa19}
\end{align}
where $S^{priv}= \frac{1}{n}Z^T Z+\frac{1}{n}w_1$, and we generate $C^{*,priv} \sim N(0,\hat{\sigma}^{2,priv} S^{priv})$.  $w_1,w_2$ are corresponding Laplacian noises, and $w_1^{*},w_2^{*}$ are the Laplacian noises generated by the same distribution of $w_1,w_2$ to account for the addition randomness in privacy protection. With $\hat{\beta}^{priv}$ and $\hat{\beta}^{*,priv}$, we can proceed following Algorithm \ref{alg2}. 

\section{Bootstrap Theory}

While the usual bootstrap estimate $\hat{\beta}_{\max}^{*,priv}$ fails to address selection bias under DP, Theorem 2 states that with a correction term and if $0<r<0.5$, our proposed method in Algorithm \ref{alg1} can deliver valid lower confidence limit for $\beta_{\max}$ and the lower confidence limit is efficient as it achieves the nominal level as $n$ goes to infinite.

\begin{theorem} For any tuning parameter $0<r<0.5$, we have that $\hat{\beta}_{\max,modified}^{*,priv}$ is consistent:  
\begin{align}
\setlength{\abovedisplayskip}{3pt}
\setlength{\belowdisplayskip}{3pt}
    &\quad \sup\limits_{x\in R} |P^{*}(\sqrt{n}(\beta_{\max,modified}^{*}-\hat{\beta}_{\max}^{priv})\leq x) \nonumber\\
    &\quad -P(\sqrt{n}(\hat{\beta}_{\max}^{priv}-\beta_{\max})\leq x)| \rightarrow 0 \nonumber 
\end{align}
\end{theorem} 
$\quad$as $n \rightarrow \infty$, in probability w.r.t $P$.

As for the partially privatized cases, the consistent property still holds by replacing $\hat{\alpha}^{priv}$ with $\hat{\alpha}=(\hat{\beta}^{priv},\gamma)$ in the proof, and the details are contained in Appendix. The justification of cross-validation and bias-reduced estimate can also be found in Appendix.

\section{Experiments}
In this section, we take multivariate Gaussian case as an example and conduct Monte Carlo simulation to demonstrate the benefits of the proposed method. Results of other scenarios can be found in the Appendix. 

We consider a simple setting with data $\mathbf{x_1},\cdots,\mathbf{x_n} \sim N(\mu,\Sigma)$ where  $\Sigma$ is identity matrix. The parameter of interest is $\mu$. We generate random samples of size $n = 800$ and use 1000 Monte Carlo samples in evaluating the empirical coverage and average distance from the true maximum value and the estimated $95\%$ lower confidence limit.  We consider tuning parameter $r= 1/30,1/15,1/10,1/5$, and the tuning parameter chosen by cross-validation.

For comparison, we adopt the naive privatized method, where we construct the CI by normal approximation with the estimated privatized extrema and its standard error considered in \cite{guo2021inference}. We also adopt the Bonferroni method to make comparison. The non-private naive method has the same structure except Laplacian noise. We also compare a semi-naive bootstrap method by setting tuning parameter $r=0.5$, which implies that we do not add bias correction term in the bootstrap.

For simplification, we use the following abbreviations: (1) \textbf{PPB}: privatized parameter bootstrap; (2) \textbf{NPB}: non-private praramer bootstrap; (3) \textbf{ParPB}: partially privatized parameter bootstrap; (4) \textbf{rPPB}: privatized parameter bootstrap adding no corresponding Laplacian noise to account for randomness. We consider different scenarios to demonstrate the benefits of the proposed method as follow.

\textbf{Bias Correction.} We set privacy budget $\varepsilon = 1.5$ as an example to see the effect of bias correction and let $\mu = (0,0)$ or $(0,1)$. From Table \ref{tb1}, we see that bias-correction plays an important role in achieving nominal coverage in both private setting and non-private setting. The proposed methods work well with cross-validation. 

\begin{table}[h]
\centering
\small
\begin{tabular}{c|c|c|c|c|c}
    \hline
    \multicolumn{2}{c|}{ true parameter }& \multicolumn{2}{c|}{$\mu = (0,0)$}  & \multicolumn{2}{c}{$\mu = (0,1)$}\\
    \hline
     \multicolumn{2}{c|}{ standard } & coverage &length &coverage & length  \\
     \hline
     \multirow{7}{3em}{\textbf{PPB}} & r = -$\infty$ & 0.939 & 0.067 & 0.952 & 0.069\\
      & r = 1/30 & 0.934 & 0.065 & 0.947 & 0.068\\
      & r = 1/15 & 0.933 & 0.065 & 0.945 & 0.068\\
      & r = 1/10 & 0.932 & 0.065 & 0.943 & 0.068\\
      & r = 1/5 & 0.924 & 0.063 & 0.939 & 0.066\\
      & r = 0.5 & 0.889 & 0.063 & 0.910 & 0.058\\
      & cv & 0.934 & 0.065 & 0.947 & 0.068\\
      \hline
      \multirow{7}{3em}{\textbf{NPB}} & r = -$\infty$ & 0.942 & 0.078 & 0.969 &0.097\\
      & r = 1/30 & 0.939 & 0.077 & 0.950 & 0.084\\
      & r = 1/15 & 0.938 & 0.076 & 0.949 & 0.083 \\
      & r = 1/10 & 0.938 & 0.076 & 0.948 & 0.082 \\
      & r = 1/5 & 0.927 & 0.074 & 0.947 & 0.081 \\
      & r = 0.5 & 0.903 & 0.055 & 0.947 & 0.081\\
      & cv & 0.937 & 0.073 & 0.947 & 0.082 \\
      \hline
      \multicolumn{2}{c|}{private naive} & 0.875 & 0.050 & 0.894 & 0.082\\
      \hline
      \multicolumn{2}{c|}{non-private naive} & 0.905 & 0.052 & 0.953 & 0.083\\
      \hline
      \multicolumn{2}{c|}{private Bonferroni} & 0.912 & 0.066 & 0.954 & 0.100  \\
      \hline
      \multicolumn{2}{c|}{\scriptsize{non-private Bonferroni}} & 0.965 & 0.071 & 0.976 & 0.099\\
      \hline
\end{tabular}
      \caption{Simulation results of Gaussian case with $k=2$}
      \label{tb1}
\end{table}

\textbf{Randomness in Bootstrap.} Following the setting in \textbf{Bias correction}, we demonstrate the importance of accounting for randomness in bootstrap by skipping the Step 6 in Algorithm \ref{alg2}. In multivariate Gaussian case, that is to say that we do not add Laplacian noise $w_1^*$ in Eq. \ref{equa8}. The results are shown in Table \ref{tb3}. We can observe that the coverage is unsatisfying. Thus it is essential to account for randomness induced by noise term in constructing CI with private data.

\begin{table}
\centering
\small
\begin{tabular}{c|c|c|c|c|c}
    \hline
    \multicolumn{2}{c|}{ true parameter }& \multicolumn{2}{c|}{$\mu = (0,0)$}  & \multicolumn{2}{c}{$\mu = (0,1)$}\\
    \hline
     \multicolumn{2}{c|}{ standard } & coverage &length &coverage & length  \\
     \hline
     \multirow{7}{3em}{\textbf{rPPB}} & r = -$\infty$ & 0.890 & 0.063 & 0.942 & 0.098\\
      & r = 1/30 & 0.882 & 0.062 & 0.916 & 0.085\\
      & r = 1/15 & 0.881 & 0.061 & 0.913 & 0.084\\
      & r = 1/10 & 0.880 & 0.061 & 0.911 & 0.083\\
      & r = 1/5 & 0.873 & 0.059 & 0.910 & 0.082\\
      & r = 0.5 & 0.844 & 0.051 & 0.909 & 0.082\\
      & cv & 0.878 & 0.061 & 0.912 & 0.083\\
      \hline
\end{tabular}
      \caption{Parametric bootstrap with no randomness of Gaussian case with $k=2$}\label{tb3}
\end{table}

\begin{figure}[ht]
\begin{minipage}[t]{0.35\linewidth}
\centering
\includegraphics[scale=0.5]{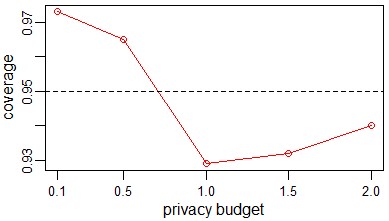}
\caption{Coverage with tuning privacy budget}
\label{fig1}
\end{minipage}%
\hfill
\begin{minipage}[t]{0.5\linewidth}
\centering
\includegraphics[scale=0.5]{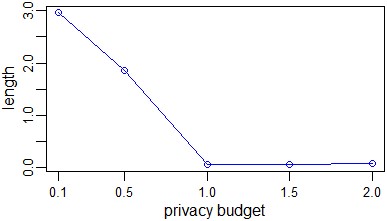}
\caption{Length with tuning privacy budget}
\end{minipage}
\label{fig2}
\end{figure}

\textbf{Privacy Budget.} We compare 5 privacy budgets $\varepsilon = 0.1,0.5,1,1.5,2$ in differential privacy with fixed tuning parameter $r = 1/10$, and the remaining settings follow \textbf{Bias correction}. The results of coverage and length are shown in Figure \ref{fig1} and Figure \ref{fig2}.

We can see that there is a trade-off between length and privacy protection. With a smaller $\varepsilon$, the privacy protection is stronger with the sacrifice in length and efficiency. With a larger $\varepsilon$, the length decreases due to less randomness induced by Laplacian noise, but the privacy protection becomes weaker. This also suggests that partially private method can help improve efficiency as long as it is relevant. The experiment results for partially privatized method is included in Appendix,  where the modification turns out to work well and leads to more efficient CI for extrema parameter.

\textbf{Dimension of Parameter.} Following the setting in \textbf{Bias correction} but with $k = 8$, Table \ref{tb2} reports the results of parametric bootstrap with tuning parameter $r = 1/10$ and $r = 0.5$ and naive methods. The privacy budget is set as $\varepsilon = 3$. More results with larger $k$ are shown in Appendix which demonstrate that when the dimension is getting higher, the advantage of our method becomes more significant. 

\begin{table}
\centering
\small
\begin{tabular}{c|c|c|c|c|c}
    \hline
    \multicolumn{2}{c|}{ true parameter }& \multicolumn{2}{c|}{$\mu = (0,\cdots,0)$}  & \multicolumn{2}{c}{$\mu = (0,\cdots,0,1)$}\\
    \hline
     \multicolumn{2}{c|}{ standard } & coverage &length &coverage & length  \\
     \hline
     \multirow{2}{3em}{\textbf{PPB}} & r = 1/10 & 0.931 & 0.058 & 0.954 & 0.069\\
      & r = 0.5 & 0.717 & 0.054 & 0.941 & 0.087\\
      \hline
      \multicolumn{2}{c|}{private naive} & 0.613 & 0.009 & 0.928 & 0.080\\
      \hline
\end{tabular}
      \caption{Simulation results of Gaussian case with $k = 8$}
      \label{tb2}
\end{table}

\section{Conclusion}
We propose a method to construct a CI for the extrema parameter under privacy scheme, which is efficient and easy to be implemented. We validate it by both analysis and experiments. Via a carefully designed privatized bootstrap procedure, selection bias in extrema parameter is appropriately adjusted under differential privacy and the randomness induced by Laplace noise is well accounted for. We also propose a partially privatized strategies which can help avoid wasting privacy budget for some application scenarios.

\section*{Acknowledgments}
This work was supported by HKUST IEG19SC04 and the Project of Hetao
Shenzhen-HKUST Innovation Cooperation Zone HZQB-KCZYB-2020083.

\bibliography{reference}
\end{document}